\documentclass[12pt,a4paper]{article}

\usepackage[utf8]{inputenc}
\usepackage[english]{babel}
\usepackage{fullpage}
\usepackage{csquotes}
\usepackage[
    backend=biber,
    style=numeric,
    block=ragged,
    url=false,
    sorting=none,
    firstinits
]{biblatex}
\usepackage{amsmath}
\usepackage{amssymb}
\usepackage{amsthm}
\usepackage{fixmath}
\usepackage{tikz}
\usepackage{tikz-inet}

\tikzset{every node/.style = {node distance=0em}}

\title{A token-passing net implementation of\\
optimal reduction with embedded read-back}
\author{Anton Salikhmetov}

\newcommand{\wait}{\text{Wait}}
\newcommand{\amb}{\text{Amb}}
\newcommand{\hold}{\text{Hold}}
\newcommand{\call}{\text{Call}}
\newcommand{\eval}{\text{Eval}}
\newcommand{\decide}{\text{Decide}}
\newcommand{\ar}{\text{Ar}}
\newcommand{\fv}{\text{FV}}

\newtheorem*{conjecture}{Conjecture}

\begin{document}
\maketitle

\begin{abstract}
In this paper, we introduce a new interaction net implementation of optimal reduction for pure untyped lambda calculus.
Unlike others, our implementation allows to reach normal form regardless of interaction net reduction strategy using the approach of so-called token-passing nets.
Another new feature is the read-back mechanism also implemented without leaving the formalism of interaction nets.
\end{abstract}

\section{Introduction}

Optimal reduction was first formalized by Lev\'y in 1980~\cite{levy} and implemented by Lamping in 1989~\cite{lamping}.
Since then, the special case of graph rewriting systems used by Lamping was distinguished as interaction nets and analyzed further by Lafont~\cite{lafont}.
Also, Lamping's optimal algorithm has been redefined using the formalism of interaction nets.

A good introduction into interaction nets can be found in \cite{tutorial} by Fern\'andez, and the problem of optimal reduction is covered by Asperti and Guerrini~\cite{optimal} in great detail.

Recently, interaction nets were described as capable to encode reduction strategies directly~\cite{strategies}.
Also, a whole new approach of token-passing nets was introduced and demonstrated to implement call-by-value and call-by-name evaluation of $\lambda$-terms using pure interaction nets thanks to Sinot~\cite{sinot1}.
Finally, interaction nets with non-deterministic extension allowed him to achieve token-passing net implementation of call-by-need evaluation~\cite{sinot2}.
Almeida, Pinto, and Vilaça analyzed and generalized this approach to a wider class of systems~\cite{tokenpassing}.
They also stated a (currently still open) question whether Mackie's closed reduction~\cite{closed} could be implemented using token-passing nets.

To the best of our knowledge, optimal reduction has not yet been implemented using the approach of token-passing until this paper.
We also found a way to implement read-back using the same technique.

In this paper, we will define an interaction system for optimal reduction of pure untyped $\lambda K$-terms with two features: reaching normal form (if any) regardless of interaction net reduction strategy and producing textual representation of the normal form without leaving the formalism of interaction nets.
The first feature is in contrast with other implementations that require avoiding interactions in disconnected parts of a net.
The second feature makes more sophisticated use of \textit{eval} and \textit{return} tokens which Sinot's token-passing nets are based on.

\section{Non-deterministic extension}

We work in interaction calculus~\cite{calculus} extended with a non-deterministic agent $\amb$~\cite{amb}. We represent this agent in a more conservative fashion than it was suggested in the original paper. Specifically, we prepend the list of auxiliary ports of $\amb$ with its extra principal port and introduce the following conversion:
$$
\langle\vec t\ |\ t = \amb(u, v, w), \Delta\rangle 
=
\langle\vec t\ |\ u = \amb(t, v, w), \Delta\rangle.
$$

We assume that any interaction system's signature $\Sigma$ is implicitly extended by $\amb$ with ${\ar(\amb) = 3}$, while its set of rules is implicitly extended with
$$
\forall \alpha \in \Sigma:
\alpha[\vec x] \bowtie \amb[y, \alpha(\vec x), y].
$$

Graphically, $\amb$ can be represented as follows:
$$
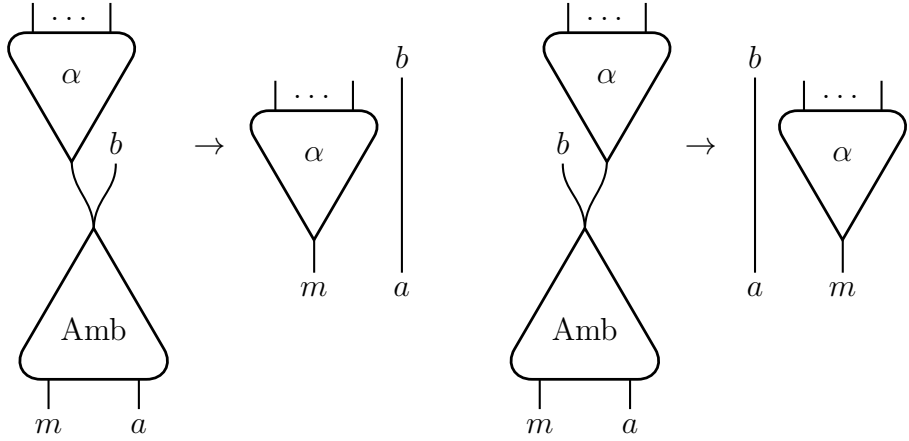
\begin{tikzpicture}[baseline=(b.base)]
\matrix[row sep=2em]{
\inetcell[opacity=0](p){$X$}[U] \\
\inetcell(a){$\amb$}[U] \\ };
\node (m) [below=of a.above right pax] {$a$};
\node (t) [below=of a.above left pax] {$m$};
\node (b) [above=of p.right pax] {$b$};
\inetcell[above=of p.left pax](x){$\phantom x\alpha\phantom x$}[D]
\node (d) [above=of x.middle pax] {$\dots$};
\inetwirefree(a.left pax)
\inetwirefree(a.right pax)
\inetwirefree(x.left pax)
\inetwirefree(x.right pax)
\inetwire(a.pal)(p.right pax)
\inetwire(a.pal)(x.pal)
\end{tikzpicture}
\rightarrow
\begin{tikzpicture}[baseline=(x)]
\matrix[row sep=1em]{
& \node (b) {$b$}; \\
\inetcell(x){$\phantom x\alpha\phantom x$}[D] & \\
\node (m) {$m$}; & \node (t) {$a$}; \\ };
\node (d) [above=of x.middle pax] {$\dots$};
\inetwirefree(x.left pax)
\inetwirefree(x.right pax)
\inetwirecoords(x.pal)(m)
\inetwirecoords(t)(b)
\end{tikzpicture}
\qquad
\begin{tikzpicture}[baseline=(b.base)]
\matrix[row sep=2em]{
\inetcell[opacity=0](p){$X$}[U] \\
\inetcell(a){$\amb$}[U] \\ };
\node (m) [below=of a.above right pax] {$a$};
\node (t) [below=of a.above left pax] {$m$};
\node (b) [above=of p.left pax] {$b$};
\inetcell[above=of p.right pax](x){$\phantom x\alpha\phantom x$}[D]
\node (d) [above=of x.middle pax] {$\dots$};
\inetwirefree(a.left pax)
\inetwirefree(a.right pax)
\inetwirefree(x.left pax)
\inetwirefree(x.right pax)
\inetwire(a.pal)(p.left pax)
\inetwire(a.pal)(x.pal)
\end{tikzpicture}
\rightarrow
\begin{tikzpicture}[baseline=(x)]
\matrix[row sep=1em]{
\node (b) {$b$}; & \\
& \inetcell(x){$\phantom x\alpha\phantom x$}[D] \\
\node (t) {$a$}; & \node (m) {$m$}; \\ };
\node (d) [above=of x.middle pax] {$\dots$};
\inetwirefree(x.left pax)
\inetwirefree(x.right pax)
\inetwirecoords(x.pal)(m)
\inetwirecoords(t)(b)
\end{tikzpicture}
$$

\section{Optimal reduction}

The pure interaction net implementation of optimal reduction upon which we base our interaction system is the main one used through most the book by Asperti and Guerrini; see interaction rules on~\cite[p.~40]{optimal} and initial encoding on~\cite[p.~41]{optimal}.
We preserve both original interaction rules and original initial encoding with exceptions for the $\beta$-reduction $\lambda_i \bowtie @_i$ rule modified and free variables allowed in a given $\lambda$-term.
The former modification is to be covered in the section about a waiting construct, and the latter feature will be a part of a read-back mechanism embedded into our interaction system.
Together, they constitute our token-passing net implementation of optimal reduction.

In order to make it easier to work with interaction calculus below, we will denote the (graphical) encoding $[M]$ of a given $\lambda$-term $M$ as (textual) configuration instead:
$$
\langle x\ |\ [M, x]\rangle,
$$
where $x$ is a name for the only free port in its interface, and $[M, x]$ is a multiset of equations that correspond to the initial encoding $[M]$.

\section{Waiting construct}

The main modification to the optimal algorithm on the way to its token-passing net version is to replace the $@_i[x, y] \bowtie \lambda_i[x, y]$ rule with
$$
@_i[x, y] \bowtie \lambda_i[\wait(z, \hold(z, x)), y],
$$
extending original signature $\Sigma_O$ with a set of agents that are used to represent what we call a waiting construct:
$$
\Sigma_W = \Sigma_O \cup \{\call, \eval, \wait, \hold, \decide\},
$$
with $\ar(\call) = 0$, $\ar(\eval) = 1$, and $\ar(\wait) = \ar(\hold) = \ar(\decide) = 2$.

The waiting construct is meant to propagate through the body of abstraction after substitution, blocking possibly unnecessary $\beta$-redexes until they are called.
This mechanism consists of several additional interaction rules we define through the rest of this section.
Since the process of deciding whether a given redex is needed has non-deterministic nature, it is the waiting construct that requires non-deterministic extension for interaction nets we discussed earlier.

In particular, interaction between a fan-in agent (denoted below as $\delta_i$) and $\wait$ results in creation of an ambiguous $\decide$ agent with two principle ports.
The latter one is simulated using $\amb$ as follows:
$$
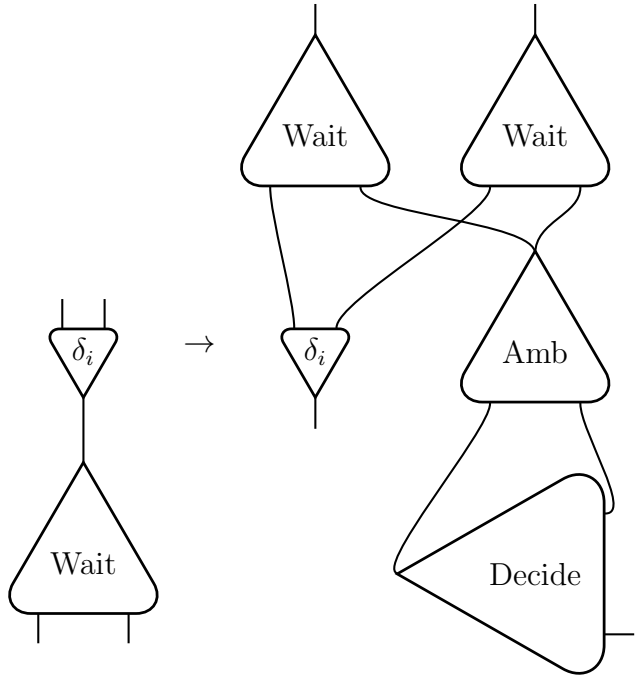
\begin{tikzpicture}[baseline=(agent)]
\matrix[row sep=2em]{
\inetcell(agent){$\delta_i$}[D] \\
\inetcell(start){$\wait$}[U] \\ };
\inetwirefree(agent.left pax)
\inetwirefree(agent.right pax)
\inetwirefree(start.left pax)
\inetwirefree(start.right pax)
\inetwire(agent.pal)(start.pal)
\end{tikzpicture}
\rightarrow
\begin{tikzpicture}[baseline=(copy)]
\matrix[row sep=2em]{
\inetcell(start1){$\wait$}[U] &
\inetcell(start2){$\wait$}[U] \\
\inetcell(copy){$\delta_i$}[D] &
\inetcell(amb){$\amb$}[U] \\
& \inetcell(wait){$\decide$}[L] \\ };
\inetwirefree(start1.pal)
\inetwirefree(start2.pal)
\inetwire(start1.left pax)(copy.right pax)
\inetwire(start2.left pax)(copy.left pax)
\inetwire(amb.pal)(start1.right pax)
\inetwire(amb.pal)(start2.right pax)
\inetwirefree(copy.pal)
\inetwirefree(wait.left pax)
\inetwire(wait.pal)(amb.left pax)
\inetwire(wait.right pax)(amb.right pax)
\end{tikzpicture}
$$

Unblocking evaluation happens through $\wait \bowtie \eval$ interaction which is implemented in a fashion similar to how evaluation strategies are encoded in Mackie's paper.
In our interaction system, the corresponding interaction rules are different due to the simulation of multiple principle ports mentioned above:
$$
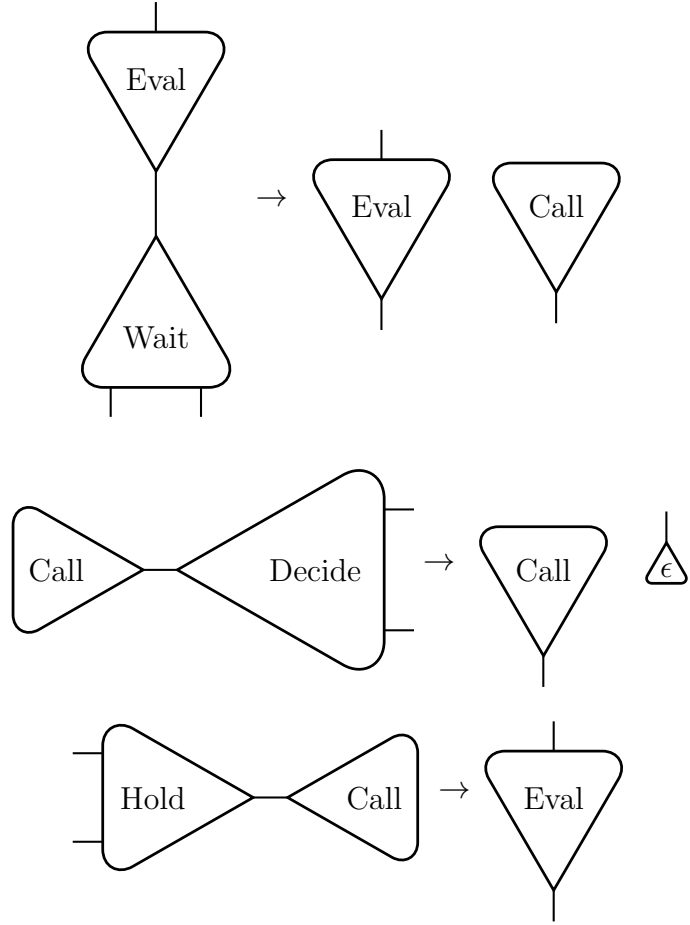
\begin{tikzpicture}[baseline=(start.above pal)]
\matrix[row sep=2em]{
\inetcell(agent){$\eval$}[D] \\
\inetcell(start){$\wait$}[U] \\ };
\inetwirefree(agent.middle pax)
\inetwirefree(start.left pax)
\inetwirefree(start.right pax)
\inetwire(agent.pal)(start.pal)
\end{tikzpicture}
\rightarrow
\begin{tikzpicture}[baseline=(agent)]
\matrix[row sep=2em, column sep=1em]{
\inetcell(agent){$\eval$}[D] &
\inetcell(go){$\call$}[D] \\ };
\inetwirefree(agent.middle pax)
\inetwirefree(agent.pal)
\inetwirefree(go.pal)
\end{tikzpicture}
$$
$$
\begin{tikzpicture}[baseline=(go)]
\matrix[row sep=2em, column sep=1em]{
\inetcell(go){$\call$}[R] &
\inetcell(wait){$\decide$}[L] \\ };
\inetwirefree(wait.left pax)
\inetwirefree(wait.right pax)
\inetwire(wait.pal)(go.pal)
\end{tikzpicture}
\rightarrow
\begin{tikzpicture}[baseline=(go)]
\matrix[row sep=2em, column sep=1em]{
\inetcell(go){$\call$}[D] &
\inetcell(wait){$\epsilon$}[U] \\ };
\inetwirefree(go.pal)
\inetwirefree(wait.pal)
\end{tikzpicture}
$$
$$
\begin{tikzpicture}[baseline=(wait)]
\matrix[row sep=2em, column sep=1em]{
\inetcell(wait){$\hold$}[R] &
\inetcell(go){$\call$}[L] \\ };
\inetwirefree(wait.left pax)
\inetwirefree(wait.right pax)
\inetwire(wait.pal)(go.pal)
\end{tikzpicture}
\rightarrow
\begin{tikzpicture}[baseline=(eval)]
\inetcell(eval){$\eval$}[D]
\inetwirefree(eval.pal)
\inetwirefree(eval.middle pax)
\end{tikzpicture}
$$

The rest of propagation and garbage collection interaction rules are more or less straightforward, so we use Lafont's notation instead to put them all together:
\begin{align*}
\eval[\lambda_i(x, y)] &\bowtie \lambda_i[x, \eval(y)]; \\
\eval[\delta_i(x, y)] &\bowtie \delta_i[x, y]; \\
\eval[x] &\bowtie \wait[\eval(x), \call]; \\
\call &\bowtie \hold[x, \eval(x)]; \\
\delta_i[\wait(x, \amb(y, \decide(z, v), v)), \wait(w, y)] &\bowtie \wait[\delta_i(x, w), z]; \\
\call &\bowtie \decide[\call, \epsilon]; \\
\epsilon &\bowtie \decide[x, x]; \\
@_i[x, \wait(y, \hold(@_i(x, y), \wait(v, w)))] &\bowtie \wait[v, w]; \\
\alpha[\wait(x, y)] &\bowtie \wait[\alpha(x), y],
\end{align*}
where $\alpha$ is a bracket or a croissant.
The only interaction rule that is worth a comment is $@_i \bowtie \wait$, because rather than propagating through application, the waiting construct has to initiate another waiting construct on the way to the root of application. Otherwise, in general case an $\epsilon$ agent that performs garbage collection from the root of application may be unable to reach either of application's sides.
In that case, it could result in leaving a non-interacting disconnected net which is still blocked garbage we would like to avoid.

\section{Read-back}

We denote the set of $\lambda$-terms as $\Lambda$, and $C[\phantom M]$ means a context~\cite{lambda}, i.~e.~a $\lambda$-term with one hole, while $C[M]$ is the result of placing $M$ in the hole of the context $C[\phantom M]$.

In order to embed read-back mechanism into our interaction system, we further extend its signature with two additional kinds of agents:
$$
\Sigma = \Sigma_W
\cup
\{\top\}
\cup
\{a_M\ |\ M \in \Lambda\}
\cup
\{r_{C[\phantom M]}\ |\ \text{$C[\phantom M]$ is a context}\},
$$
with ${\ar(a_M) = 0}$ and ${\ar(r_{C[\phantom M]}) = \ar(\top) = 1}$,
the \textit{atom} agent $a_M$ encoding the textual representation of a $\lambda$-term $M$ and the \textit{read} agent $r_{C[\phantom M]}$ performing read-back in the context of $C[\phantom M]$.
In particular, agents $a_M$ make it possible to represent free variables in a given $\lambda$ term being encoded into interaction net.

While encoding $\lambda$-terms into our interaction system, we will distinguish their free variables from their bound variables.
So, let us mark all free variables in a $\lambda$-term $M$ using the following operation: ${M^\bullet \equiv M[\vec x := \vec x^\bullet]}$, where ${(\vec x) = \fv(M)}$.
Then, $\lambda$-term $M$ can be mapped to configuration
$$
\langle x\ |\ \eval(r_{[\phantom M]}(\top(x))) = y, [M^\bullet, y]\rangle,
$$
where original encoding $[M, x]$ is extended with $[x^\bullet, y] = \{a_x = y\}$.

Our read-back mechanism mainly consists of the following three interaction rules.
$$
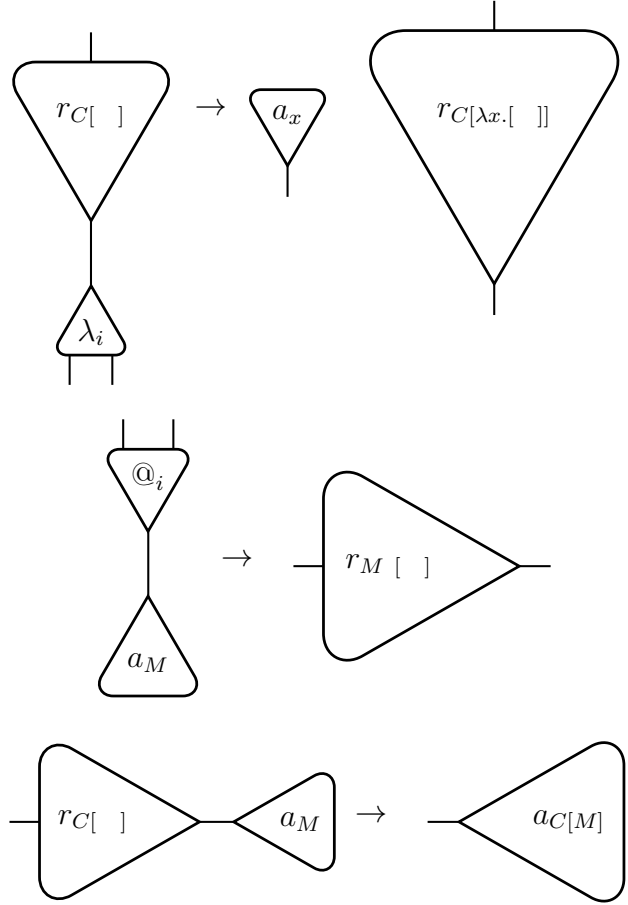
\begin{tikzpicture}[baseline=(read)]
\matrix[row sep=2em]{
\inetcell(read){$r_{C[\phantom M]}$}[D] \\
\inetcell(lambda){$\lambda_i$}[U] \\ };
\inetwirefree(read.middle pax)
\inetwirefree(lambda.left pax)
\inetwirefree(lambda.right pax)
\inetwire(read.pal)(lambda.pal)
\end{tikzpicture}
\rightarrow
\begin{tikzpicture}[baseline=(read)]
\matrix[row sep=2em, column sep=1em]{
\inetcell(atom){$a_x$}[D] &
\inetcell(read){$r_{C[\lambda x.[\phantom M]]}$}[D] \\ };
\inetwirefree(atom.pal)
\inetwirefree(read.middle pax)
\inetwirefree(read.pal)
\end{tikzpicture}
$$
$$
\begin{tikzpicture}[baseline=(atom.above pal)]
\matrix[row sep=2em]{
\inetcell(appl){$@_i$}[D] \\
\inetcell(atom){$a_M$}[U] \\ };
\inetwirefree(appl.left pax)
\inetwirefree(appl.right pax)
\inetwire(appl.pal)(atom.pal)
\end{tikzpicture}
\rightarrow
\quad
\begin{tikzpicture}[baseline=(read)]
\inetcell(read){$r_{M\ [\phantom M]}$}[R]
\inetwirefree(read.middle pax)
\inetwirefree(read.pal)
\end{tikzpicture}
$$
$$
\begin{tikzpicture}[baseline=(read)]
\matrix[row sep=2em, column sep=1em]{
\inetcell(read){$r_{C[\phantom M]}$}[R] &
\inetcell(atom){$a_M$}[L] \\ };
\inetwirefree(read.middle pax)
\inetwire(read.pal)(atom.pal)
\end{tikzpicture}
\rightarrow
\quad
\begin{tikzpicture}[baseline=(atom)]
\inetcell(atom){$a_{C[M]}$}[L]
\inetwirefree(atom.pal)
\end{tikzpicture}
$$

More formally in interaction calculus, the rules related to read-back are as follows:
\begin{align*}
r_{C[\phantom M]}[x] &\bowtie \lambda[a_y, r_{C[\lambda y.[\phantom M]]}(x)], \quad \text{where $y$ is a fresh variable}; \\
@_i[r_{M\ [\phantom M]}(x), x] &\bowtie a_M; \\
r_{C[\phantom M]}[a_{C[M]}] &\bowtie a_M; \\
r_{C[\phantom M]}[\alpha(x)] &\bowtie \alpha[r_{C[\phantom M]}(x)], \quad \text{where $\alpha$ is a bracket or a croissant}; \\
r_{C[\phantom M]}[\wait(x, y)] &\bowtie \wait[r_{C[\phantom M]}(x), y]; \\
\eval[a_M] &\bowtie a_M; \\
\alpha[a_M] &\bowtie a_M, \quad \text{where $\alpha$ is a bracket or a croissant}; \\
\top[a_M] &\bowtie a_M; \\
\top[x] &\bowtie \alpha[\top(x)], \quad \text{where $\alpha$ is a bracket or a croissant}.
\end{align*}

Now, we believe that the following statement holds true.
\begin{conjecture}
$\langle x\ |\ \eval(r_{[\phantom M]}(\top(x))) = y, [M^\bullet, y]\rangle \downarrow \langle a_N\ |\ \varnothing\rangle$
iff $N$ is normal form of $M$.
\end{conjecture}
That is, the interaction net that encodes a $\lambda$-term $M$ in our interaction system reduces to normal form (if any) with no garbage and only one agent $a_N$ in its interface, $N$ representing normal form of the encoded $\lambda$-term $M$.

\section{Further work}

This is currently ongoing work in the context of our Macro Lambda Calculus project aiming to implement pure untyped $\lambda K$-calculus using a domain-specific programming language based on interaction calculus.
Our implementation of interaction calculus has non-deterministic extension and also allows side effects.
Another significant difference from interaction calculus is lack of notion of an interface which leads to inability of following any specific strategy.
The latest stable version of our implementation is available online at \url{https://codedot.github.io/lambda/} and works in modern Web browsers with no server side, computation being performed solely on the client side.

We found that approach of token-passing nets fits this model perfectly (in particular, we have successfully implemented Sinot's call-by-need for free~\cite{mlc}).
However, the optimal algorithm and many other interaction net implementations of $\lambda$-calculus rely upon weak strategies aiming to reach interface-normal form of configurations and require external garbage collection which we would like to delegate to interaction nets instead.

Our token-passing implementation of optimal reduction with embedded read-back mechanism is a preliminary result we still consider worth sharing.
The conjecture given above is still missing a formal proof at this point.
However, no counterexamples have been found while experimenting with software implementation of these ideas.

Also, it is interesting to consider more sophisticated use of the waiting construct being propagated through the whole body of every abstraction applied to an argument, especially for run-time optimization of oracle nodes as they are known to be the most significant show-stopper for implementations of optimal reduction.

\printbibliography
\end{document}